\documentclass{WileyMSP-template}
\usepackage{braket}
\usepackage{ragged2e}
\usepackage{subcaption}
\usepackage{amsmath}
\usepackage{textcomp}

\begin{document}

\pagestyle{fancy}
\rhead{\includegraphics[width=2.5cm]{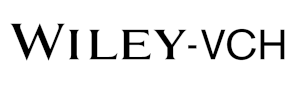}}

\title{First-Principles Investigation of Surface-Induced Effects on the Properties of Divacancy Qubits in 3C-SiC}

\maketitle

\author{Rosario G. Viglione},
\author{Giovanni Castorina},
\author{Gaetano Calogero},
\author{Giuseppe Fisicaro},
\author{Damiano Ricciarelli},
\author{Ioannis Deretzis} and
\author{Antonino La Magna}*


\begin{affiliations}
Address\\ Consiglio Nazionale delle Ricerche, Istituto per la Microelettronica e Microsistemi (CNR-IMM), Z.I. VIII Strada 5, 95121 Catania, Italy\\
Email Address: Antonino.LaMagna@imm.cnr.it
Antonino La Magna \\

\end{affiliations}




\begin{abstract}
\justifying
Neutral silicon–carbon divacancy (V$_\mathrm{Si}$V$_\mathrm{C}$) in cubic silicon carbide (3C–SiC) is a promising class of point defects for quantum technologies based on active crystalline centers. Within the theoretical framework of spin-polarized Density Functional Theory (DFT), this study examines the structural and electronic characteristics of V$_\mathrm{Si}$V$_\mathrm{C}$ centers near a hydrogen-terminated Si-rich (001) surface. A (2$\times$1):H reconstructed slab of 628 atoms represents the near-surface environment, with divacancies located at depths ranging from 0.6 to 1.2 nm in basal and axial orientations. The optimized geometries show localized relaxations, and the electronic structure reveals in-gap defect levels in both spin channels. Furthermore, examination of the zero-field splitting (ZFS) tensor demonstrates sensitivity to the orientation of the spin defects and their distance from the surface. The findings of this investigation suggest that surface proximity exerts a substantial influence on the spin Hamiltonian of divacancies, providing insight for the engineering of SiC-based qubits and nanoscale quantum devices.
\end{abstract}


\section{Introduction}
\justifying
Spin defects in wide-band gap semiconductors have emerged as essential components for quantum technologies \cite{LarsJaeger}, such as quantum sensing, computing and single-photon emission. In particular, silicon carbide (SiC) has become a key material platform due to its advanced fabrication technology, excellent thermal properties, and capacity to support various optically addressable point defects.\cite{Awschalom2018} The considerable interest for SiC materials in the field of quantum technologies is essentially due to its color centers, which possess states suitable for quantum applications.\cite{Castelletto2020-qd} The coherence properties of the color center states in various SiC polytypes at room temperature have been extensively investigated both theoretically and experimentally, using a range of techniques, that have yielded promising results.\cite{Seo2016-vy, Fazio2024-il} The neutral silicon-carbon divacancy in the cubic 3C-SiC polytype is a significant qubit candidate,\cite{Falk2014} showing a spin triplet ground state (S=1) and long spin coherence times that remain stable at room temperature,\cite{Koehl2011} similar to the nitrogen-vacancy center in diamond.\cite{Xuan2025} 

Despite the extensive examination of the general properties of SiC divacancies, their behavior near a surface, which is a critical aspect for integrating these centers into hybrid quantum devices, remains less explored.\cite{Zhu2023} In the context of device applications, defect qubits near surfaces allow for efficient coupling with photonic\cite{Lukin2023} or plasmonic structures.\cite{JiYangZhou2023} However, the surface exhibits a significant disturbance: it disrupts the symmetry of the bulk crystal, it generates localized strain fields, and may alter the local electrostatic conditions. Therefore, it is important to comprehend the interplay among surface termination, strain and defect states. 
A crucial descriptor of a spin-triplet system is the zero-field splitting (ZFS) tensor, which is defined by the axial parameter ($D$) and the transverse parameter ($E$). In bulk 3C-SiC, the V$_\mathrm{Si}$V$_\mathrm{C}$ center exhibits $C_{3v}$ symmetry, leading to  an almost zero transverse ZFS component. In the vicinity of surfaces, symmetry reduction gives rise to non-zero $E$ values, which can affect qubit performance. 
This study uses first-principles DFT simulations to investigate the impact of a hydrogen-terminated Si-rich (001) surface on the structure and spin characteristics of near-surface neutral divacancies in 3C–SiC. Our examination includes atomic relaxation, electronic defect states and ZFS tensor components, with a focus on their variation as a function of divacancy depth and orientation. Moreover, we examine the feasibility of employing 3C-SiC based qubits utilizing divacancy color centers through detailed simulations of the system’s spin states dynamics. The effects of decoherence and the possibility of preparing target states via radiofrequency (RF) pulses are investigated through these simulations.\\
The present work also relates to recent theoretical efforts aimed at modeling spin and nuclear interactions mediated by point defects in semiconductors, including studies exploring defect-assisted control of nuclear spin order within solid-state quantum architectures.\cite{Calogero2025} Such approaches provide a broader framework for understanding how microscopic defect environments influence coherence and spin–lattice coupling in SiC and related materials.

\section{Methods}
\justifying
\subsection{Spin Hamiltonian and rationale}
\justifying
Ab initio methods that rely on the Density Functional Theory are essential to calculate the properties of materials, including their structural, electronic and magnetic characteristics. These techniques are particularly useful for calibrating the parameters of electron paramagnetic resonance (EPR) Hamiltonians, which define the intricate interactions of electron spins in their surroundings. Their application is particularly important when the study focuses on defects residing at interfaces or at the proximity of surfaces, especially when relevant experimental data are either ambiguous or even absent. The total Hamiltonian for the examined spin system is expressed as:

\begin{equation}
    H = H_e + \sum_i \mathbf{S} \cdot \mathbf{A}_i \cdot \mathbf{I}_i - \sum_i \gamma_{n,i} \mathbf{B} \cdot \mathbf{I}_i + \sum_{i<j} \mathbf{I}_i \cdot \mathbf{P}_{ij} \cdot \mathbf{I}_j
    \label{eq:full_hamiltonian}
\end{equation}
where the electron spin Hamiltonian, $H_e$, is given by:
\begin{equation}
    H_e = -\gamma_e \mathbf{B} \cdot \mathbf{S} + D\left(S_z^2 - \frac{1}{3}S(S+1)\right) + E(S_x^2 - S_y^2)
    \label{eq:electron_hamiltonian}
\end{equation}

The first term in Equation (1) is the electron spin Hamiltonian, which is composed of the ZFS terms and the Zeeman interaction. The last three terms are the magnetic hyperfine interactions, the nuclear Zeeman terms and the dipolar interactions between nuclear spins, respectively. The impact of the electron-ion spin interaction is critical in determining the coherence loss of the electron spin states due to dephasing. This effect can be explicitly evaluated considering the average kinetics of one electronic spin interacting with large baths of ion spins within the cluster correlation expansion approximation \cite{Seo2016-vy, Fazio2024-il}. These calculations are beyond the scope of this paper, and here the decoherence effects on the electronic spin are effectively evaluated with the Lindblad master equation (\cite{KitsonPhysRevA2024}; see also sub-section below).
The dynamics of the open system are described by the master equation (in angular frequency units):
\begin{equation}
\label{eq:Lindblad}
    \frac{d\rho}{dt} = -\mathrm{i}\,[H,\rho]
    + \frac{\gamma}{2}\!\left( S_z\,\rho\,S_z - \tfrac{1}{2}\{S_z^2,\rho\} \right),
\end{equation}
where the corresponding dephasing collapse operator is given by $\sqrt{\gamma/2}\,S_z$. 

\subsection{Computational details}
\justifying
The electronic structure of the spin defects and the relaxed geometry of the supercell were determined within the framework of spin-polarized density functional theory, using plane wave basis sets and Optimized Norm-Conserving Vanderbilt (ONCV) pseudopotentials,\cite{Hamann2013} by means of the Quantum Espresso package.\cite{Giannozzi2009} Convergence was achieved with plane-wave cutoff energies of 80~Ry for the wavefunctions and 340~Ry for the charge density. The exchange–correlation potential was treated using the generalized gradient approximation (GGA) in the Perdew–Burke–Ernzerhof (PBE) formulation.\cite{PBE1996} 

A 628-atom slab models the Si-rich (001) surface of 3C–SiC with a (2$\times$1):H reconstruction, with a vacuum spacing of approximately 20~\AA, introduced along the surface normal to avoid interactions between periodic replicas. To reproduce bulk-like conditions, the bottom carbon atoms were saturated with hydrogen atoms, and the lowest atomic layer was partially fixed during structural relaxation. The remaining atoms were fully relaxed until the residual forces were below 0.001~eV~\AA$^{-1}$, ensuring structural equilibrium.

Neutral divacancies were introduced at four different depths (from 0.6 nm to 1.2 nm), labeled as L$n$ with $n = 6, 5, 4, 3$. For each depth, we considered two symmetry orientations relative to surface normal: "axial" ($\perp$) and "basal" ($\parallel$), as illustrated in Figure 1.\\
The ZFS tensor components were derived from the spin–spin dipolar interaction following the formalism of Rayson and Briddon,\cite{Rayson2008} using the PyZFS package,\cite{pyzfs} which enables direct post-processing of wavefunctions obtained from Quantum ESPRESSO. In this case, SCF calculations were performed with PAW pseudopotentials.\cite{Paw} As a reference, we calculated the bulk ZFS values for a 3$\times$3$\times$3 supercell of 3C-SiC. This approach allows accurate evaluation of both the axial ($D$) and transverse ($E$) components of the ZFS, which are essential for characterizing the fine-structure of spin-triplet states in semiconductors.\\
Our computational methodology builds upon established first-principles frameworks for modeling spin-active defects in solid-state quantum systems.

The spin state dynamics and decoherence processes were simulated within the full Hilbert space using the Python package \textbf{QuTiP}.\cite{lambert2024qutip5quantumtoolbox} The vector basis, operators, and initial states for the divacancy electron spin-$1$ system (three-dimensional Hilbert space) were constructed accordingly. The eigenvalues of the electronic Hamiltonian $H_e$ as a function of the magnetic field $B$ were computed to obtain the energy spectrum of the electronic triplet states, corresponding to the $m = 0, \pm1$ sublevels. Subsequently, Schrödinger-type and Lindblad-type evolutions of the spin expectation values were performed using QuTiP’s internal solvers, with parameters derived from the preceding DFT simulations incorporated into the Hamiltonian. Bulk SiC divacancies can reach unusually long $T_2$ decoherence time which can be recovered by a small $\gamma$ value in Equation \ref{eq:Lindblad}. Actually, the coherence of near-surface divacancies is not known and here we use a moderate $\gamma$ value in the analysis, with $\gamma \approx 10^{-2}\,\mathrm{rad\,s^{-1}}$. Anyhow, simulation total times are of the order $\frac{10 \pi}{\Omega_R}$ (where $\Omega_R$ denotes the Rabi frequency, set to $0.3\,\mathrm{MHz}$ for these simulations) and the Schrödinger-type evolution is representative for very small $\gamma$ values. The simulations were carried out both under free evolution (for closed and open systems) and under external Rabi excitation, modeled as a square-modulated cosine driving field. The Rabi excitation frequency was determined from the energy difference between the $m = 0$ and $m = -1$ states at maximum value of B=$0.3$ T.
The interaction with the electromagnetic field is modeled by adding a time-dependent term to the Hamiltonian (in angular frequency units):
\begin{equation}
    H_{\mathrm{em}} = 2\Omega_R \cos(\omega_R t) \frac{S_{x,\mathrm{eff}}}{2},
\end{equation}
where $\omega_R$ is the Rabi excitation frequency calculated as described above, and $S_{x,\mathrm{eff}}$ is an operator that couples the lower-energy eigenstates, given by
\begin{equation}
    S_{x,\mathrm{eff}} =
    \begin{pmatrix}
       0 & 0 & 0 \\
       0 & 0 & 1 \\
       0 & 1 & 0
    \end{pmatrix}.
\end{equation}

\section{Results and discussion}
\begin{figure}[ht!]
    \centering
    \includegraphics[width=0.8\linewidth]{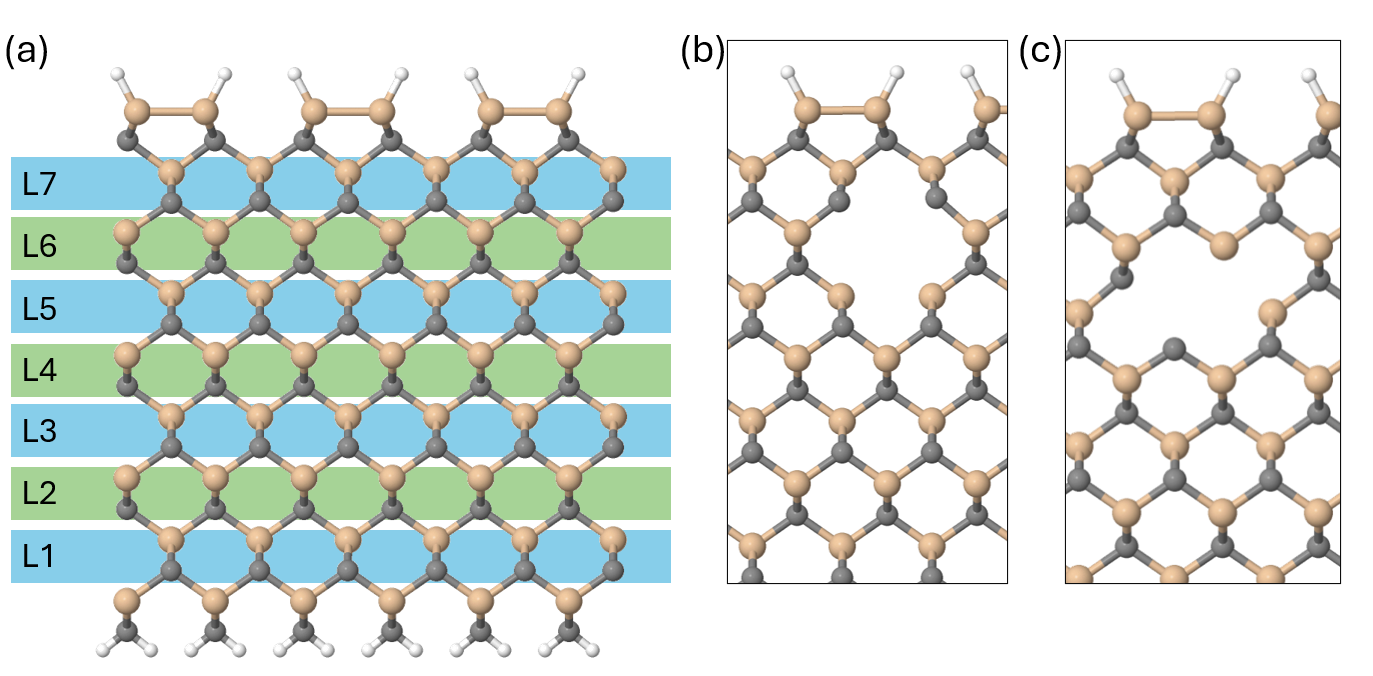}
    \caption{(a) Atomistic model of a (2$\times$1):H reconstructed 3C-SiC surface showing the various SiC dimer layers. (b) Divacancy (V$_\mathrm{Si}$V$_\mathrm{C}$) defect with an axial configuration. (c) Divacancy (V$_\mathrm{Si}$V$_\mathrm{C}$) defect with a basal configuration.}
    \label{fig:atomistic_model}
    \end{figure}

\justifying
The optimized (2$\times$1):H slab model is illustrated in Figure~\ref{fig:atomistic_model}. The passivation of the dangling bond with H is an experimentally reproducible configuration which minimizes the impact of surface reconstruction on the electronic and spin states of the defects compared to other kinds of reconstructions.\cite{Zhu2023} 
The introduction of the V$_\mathrm{Si}$V$_\mathrm{C}$ defect induces local atomic distortions, primarily within the first coordination shells. 
    

Table~\ref{tab:geometry} reports the local geometry around the divacancies. We notice that defects in close proximity to the surface (L6) exhibit a marginally more pronounced anisotropic distortion compared to defects located deeper, as the interaction with the surface becomes prominent.

\begin{figure}[h!]
    \centering
    \includegraphics[width=0.5\linewidth]{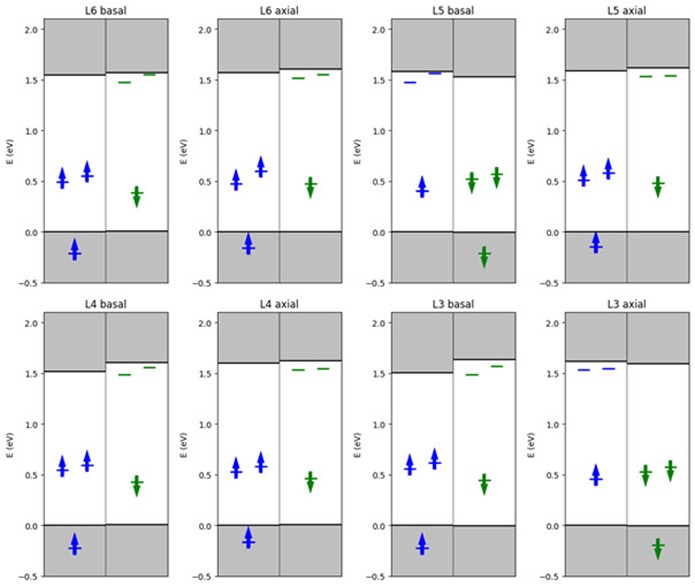}
    \caption{Defect electronic states in ($\perp$) and ($\parallel$) orientations of the  V$_\mathrm{Si}$V$_\mathrm{C}$ divacancy at different distances from the (2$\times$1):H surface. Blue and green lines indicate spin up and spin down channels, respectively, while arrows define the occupied defect states. Conduction and valence bands are shown in gray. }
    \label{fig:electronic_states}
\end{figure}

The results of spin-polarized calculations, reported in Figure~\ref{fig:electronic_states}, indicate that the ground state of the V$_\mathrm{Si}$V$_\mathrm{C}$ defect is a triplet state, for all the cases studied here. Differences in the distance of these defects from the reconstructed (2$\times$1):H surface or in their orientation relative to the surface normal direction result in small quantitative displacements (usually few millielectronvolts) of the energy levels. Such electronic robustness could be important for the potential use of these systems in quantum technology applications. Furthermore, no surface states were calculated within the band gap, due to the hydrogen passivation of the reconstructed (2$\times$1) surface. Hence, surface-induced interference effects from ordered H-passivated (2$\times$1) surfaces should be absent for V$_\mathrm{Si}$V$_\mathrm{C}$ spin qubits, indicating that a controlled generation of SiC surfaces is technologically relevant.

\begin{table}[h!]
\centering
    \caption{Local geometry around the divacancies. Bond lengths are in \AA.}
    \label{tab:geometry}
  \begin{tabular}[htbp]{@{}lll@{}}
    \hline
        Structure & C--C distances (\AA) & Si--Si distances (\AA) \\
        \hline
        Bulk & 3.309, 3.308, 3.308 & 3.100, 3.097, 3.097 \\
        \hline
        \multicolumn{3}{c}{\textbf{Axial ($\perp$) Orientation}} \\
        L6 & 3.269, 3.387, 3.382 & 3.179, 3.124, 3.077 \\
        L5 & 3.414, 3.332, 3.332 & 3.171, 3.112, 3.113 \\
        L4 & 3.382, 3.331, 3.320 & 3.176, 3.102, 3.098 \\
        L3 & 3.374, 3.324, 3.321 & 3.173, 3.104, 3.101 \\
        \hline
        \multicolumn{3}{c}{\textbf{Basal ($\parallel$) Orientation}} \\
        L6 & 3.335, 3.329, 3.330 & 3.134, 3.101, 3.100 \\
        L5 & 3.338, 3.318, 3.319 & 3.124, 3.119, 3.120 \\
        L4 & 3.341, 3.322, 3.322 & 3.130, 3.114, 3.114 \\
        L3 & 3.336, 3.323, 3.324 & 3.129, 3.108, 3.108 \\
        \hline
 \end{tabular}
 \end{table}

Table~\ref{tab:zfs} shows the computed components $D$ and $E$ of the ZFS tensor for V$_\mathrm{Si}$V$_\mathrm{C}$ defects having different distances and orientations from the (2$\times$1):H surface. Divacancies with basal ($\parallel$) orientation generally show higher values of the $D$ component with respect to both defects with axial orientation ($\perp$) and defects in bulk SiC. Hence, $D$ values could allow for the determination of V$_\mathrm{Si}$V$_\mathrm{C}$ orientations in near-surface SiC systems. On the other hand, the transversal component $E$ is sensitive to the position of the divacancy with respect to the surface. For the ($\perp$) orientation, the surface effect decreases rapidly with distance. For ($\parallel$) orientations instead, the defect induced distortions responsible for $E$ may persist more uniformly deeper into the slab before subsiding. The absence of a monotonic variation of $E$ with respect to the surface distance in this latter case may complicate the potential use of ZFS values as probes for the determination of the exact position of these defects with respect to the SiC  (2$\times$1):H surface.

\begin{table}[h!]
    \centering
    \caption{Computed components $D$ and $E$ of ZFS tensor for V$_\mathrm{Si}$V$_\mathrm{C}$ defects having different distances and orientations from the (2$\times$1):H surface.}
    \label{tab:zfs}
    \begin{tabular}[htbp]{@{}lll@{}}
        \hline
        Structure & $D$ (MHz) & $E$ (MHz) \\
        \hline
        Bulk & 1418 & 0 \\
        \hline
        \multicolumn{3}{c}{\textbf{Axial ($\perp$) Orientation}} \\
        L6 & 1384 & -166 \\
        L5 & 1369 & -57 \\
        L4 & 1395 & -39 \\
        L3 & 1401 & -30 \\
        \hline
        \multicolumn{3}{c}{\textbf{Basal ($\parallel$) Orientation}} \\
        L6 & 1457 & -45 \\
        L5 & 1470 & -62 \\
        L4 & 1466 & -64 \\
        L3 & 1466 & -63 \\
        \hline
        \multicolumn{3}{l}{Bulk (Exp.)\cite{Christle} \hspace{1cm} $\sim$1336 \hspace{1cm} } \\
        \hline
    \end{tabular}
\end{table}

\begin{figure}[h!]
  \centering
  \begin{minipage}{0.48\textwidth}
    \centering
    \includegraphics[width=\linewidth]{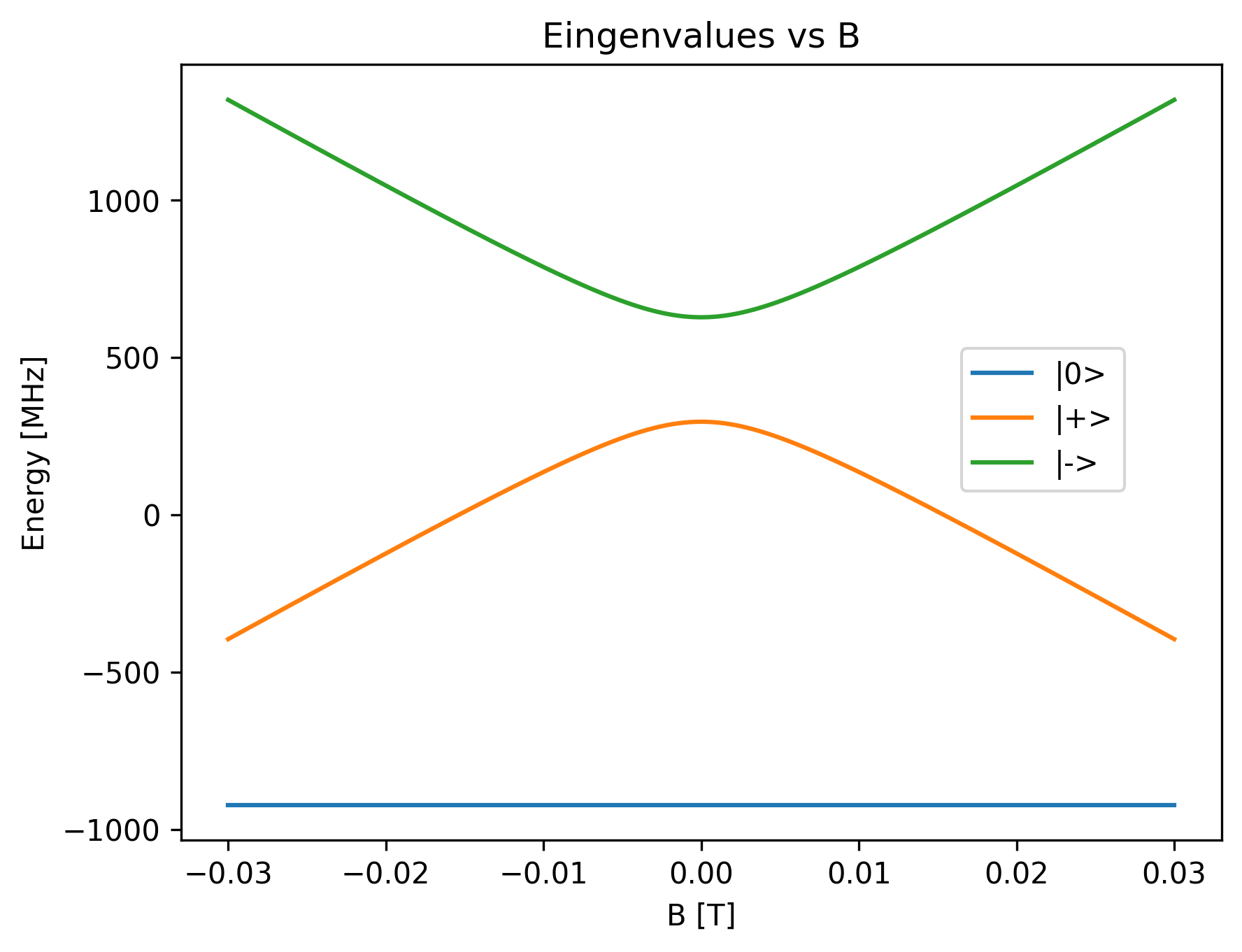}
    \subcaption*{(a)}
  \end{minipage}%
  \begin{minipage}{0.48\textwidth}
    \centering
    \includegraphics[width=\linewidth]{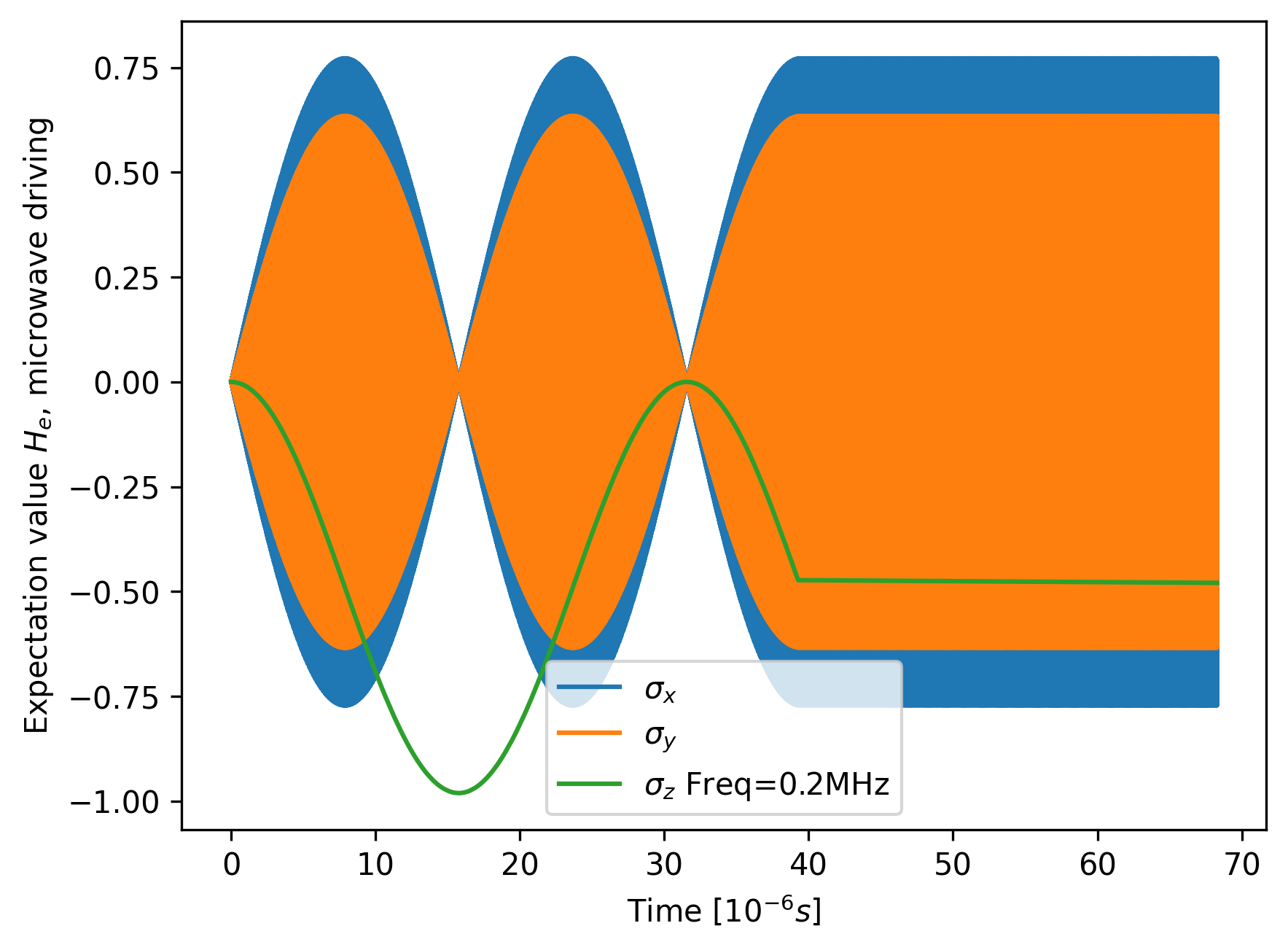}
    \subcaption*{(b)}
  \end{minipage}

  \vspace{0.3em}

  \begin{minipage}{0.48\textwidth}
    \centering
    \includegraphics[width=\linewidth]{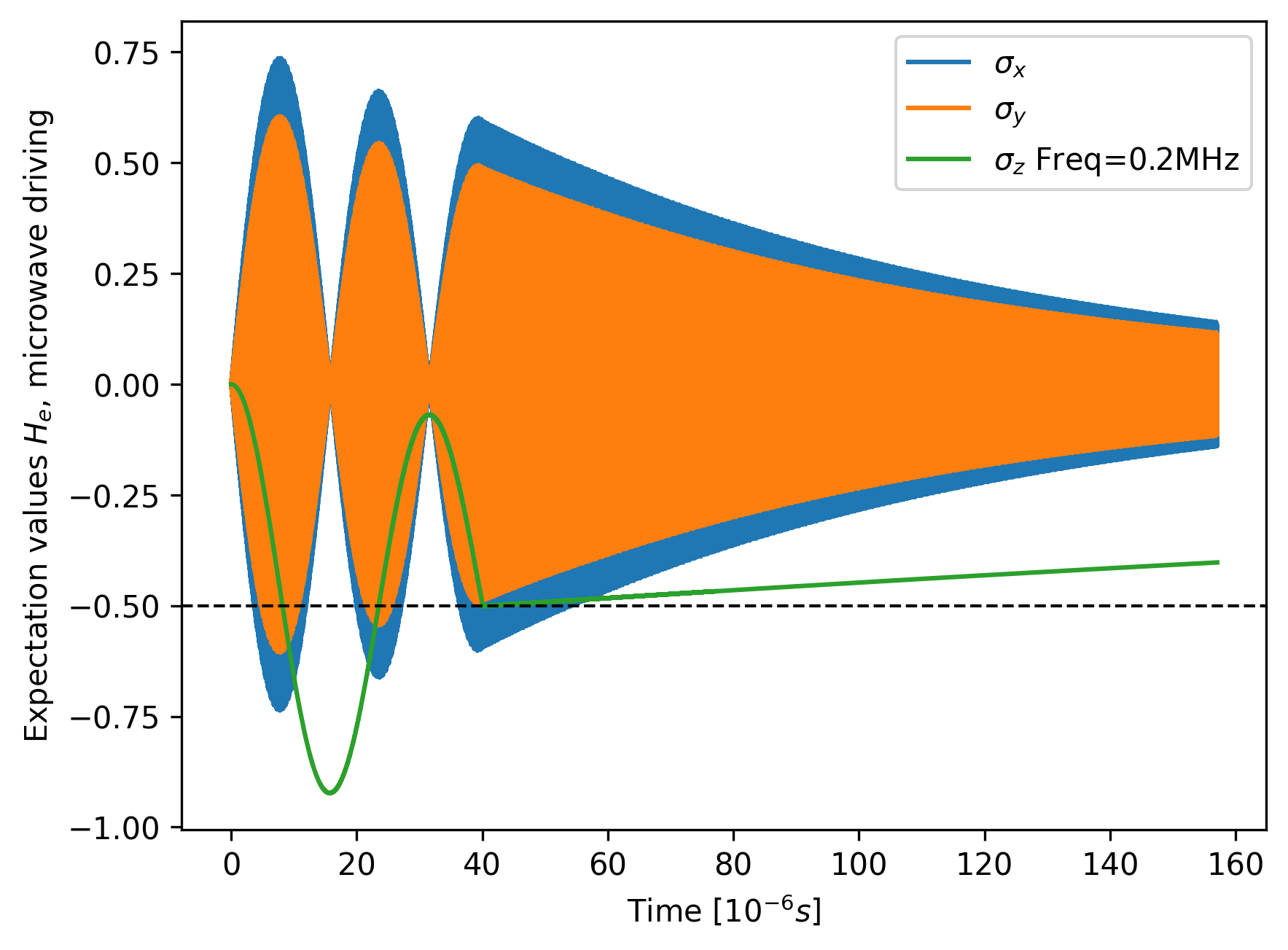}
    \subcaption*{(c)}
  \end{minipage}%
  \begin{minipage}{0.48\textwidth}
    \centering
    \includegraphics[width=\linewidth]{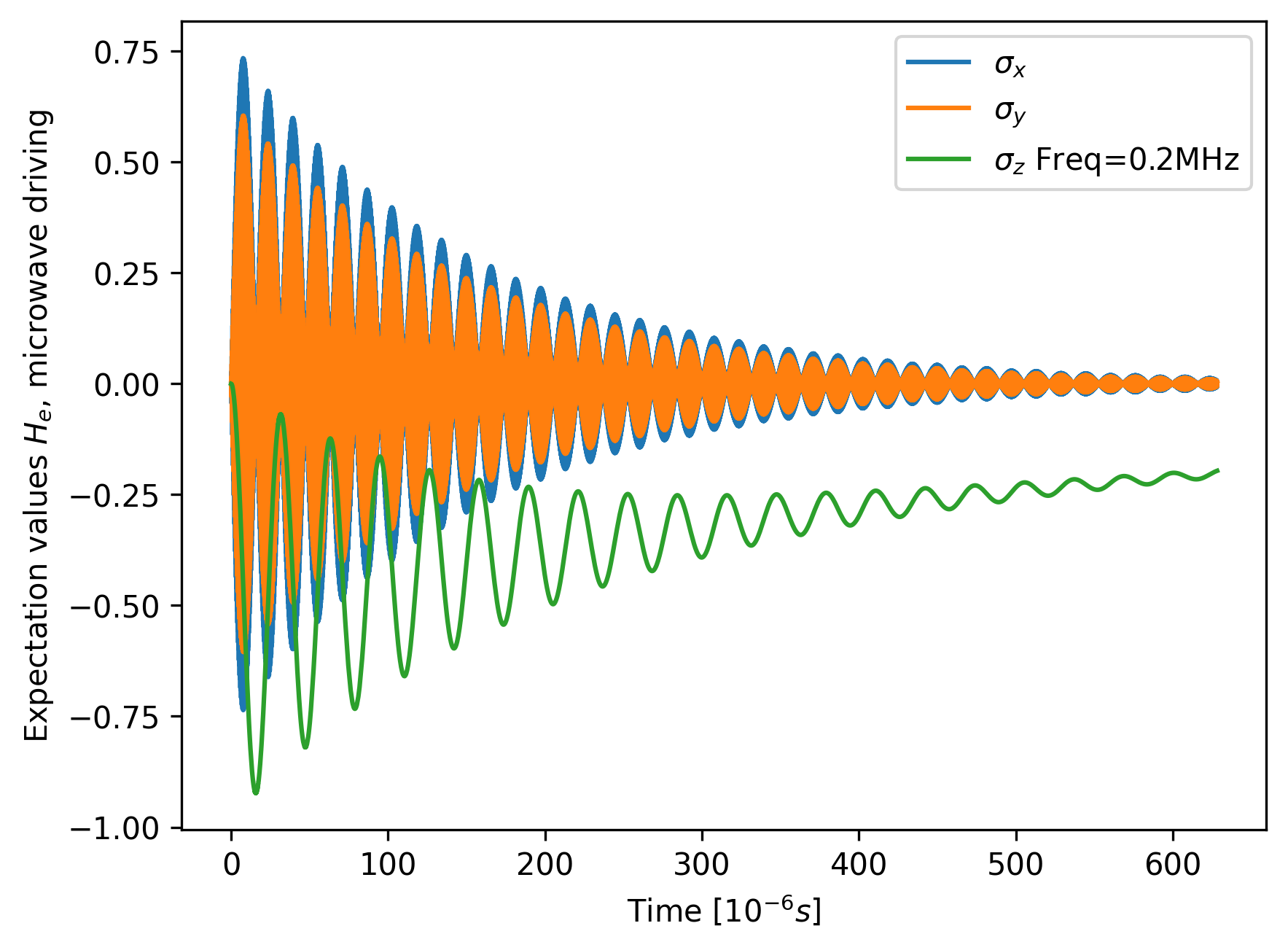}
    \subcaption*{(d)}
  \end{minipage}

  \caption{(a) Calculated eigenvalues of the model as a function of the external magnetic field $B$. The Rabi frequency is fixed at $\Omega_R = 0.2\,\text{MHz}$. (b) Time evolution of the expectation values of the spin operators under a Rabi excitation of duration $5\pi / (2\Omega_R)$ in the absence of dissipation. (c) Time evolution of the expectation values of the spin operators under a Rabi excitation of the same duration, including a dissipation term proportional to $\sigma_z$. The black line is a reference for the $\braket{\sigma_z}=-0.5$ value.(d) Time evolution of the expectation values of the spin operators under both Rabi excitation and dissipation for an extended simulation time.}
  \label{fig:libero}
\end{figure}
The \textit{ab initio} analysis demonstrates that a defect, which has a single EPR feature in symmetric 3C-SiC bulk material, results in multiplicity of spin states when its near-surface configurations are considered. These results have to be considered in the realization of manipulation protocols applied for quantum operation. Here we evaluate the basic resonant radio frequency driving used for the controlled rotation gate as the $C_nROT_e$ one.\cite{Cai2013-bc, Zaiser2016-va}    
The eigenvalues dependence and the radio frequency excitation are calculated expanding the electronic Hamiltonian only with the terms due to the external field. The model is:
\begin{equation}
    H= -\gamma_e B  S_z + D\left(S_z^2 - \frac{1}{3}S(S+1)\right) + E(S_x^2 - S_y^2) + \Omega_R \cos(\omega_R t) S_{x,\mathrm{eff}}
\end{equation}
Here, $D$ and $E$ are taken from Table~\ref{tab:zfs} for the L6 configuration (axial), and the magnetic field $B$ is varied within the range $[-0.3\,\mathrm{T},\,0.3\,\mathrm{T}]$.

Static properties of the divacancy electron spin system and time evolution during processes are investigated. The fundamental energy level structure of the divacancy electron spin in response to a static magnetic field along the z axis is shown in Figure ~\ref{fig:libero}(a), which displays the energy eigenvalues versus the external magnetic field $B$. The $\ket{0}$ state (blue line) exhibits minimal energy change, consistent with its $m_s=0$ character and weak dependence on the B field in this range. Its energy near -950MHz is primarily set by the zero-field splitting (ZFS) parameters D and E. Conversely, the $\ket{+}$ (orange) and $\ket{-}$ (green) states show significant Zeeman splitting, with their energy separation increasing approximately linearly with $|B|$. At $B=0$, these states are split only by the rhombic ZFS term (E), which causes mixing between the $m_s = \pm 1$ basis states.

The qubit initialization is obtained with off-resonance excitation, which allows the recovery of the $m_s = 0$ lowest energy state (in a wide range of $B$ values) with high fidelity. Therefore, it is interesting to examine, starting from the $\ket{0}$ state, how the system evolves and whether the maximally entangled state $\frac{1}{\sqrt{2}}(\ket{0} + \ket{-1})$ can be achieved in order to obtain a system that can be employed as a qubit when a local polarization exchange with a memory spin is allowed.\cite{Zaiser2016-va} To this end, radiofrequency pulses based on the previously calculated Rabi excitation are used to perform $\frac{\pi}{2}$ rotations of the qubit state.

Starting from a closed system, it is also insightful to analyze the effect of applying a pulse corresponding to five $\frac{\pi}{2}$ rotations (in general an odd number of $\frac{\pi}{2}$ rotations), in order to reach a mixed final state. As shown in Figure~\ref{fig:libero}(b), the system attains the desired final state in this case.

This discussion changes slightly in the case of open systems, as shown in Figure~\ref{fig:libero}(c). In this situation, the decoherence introduced by the dissipative terms ($\gamma = 5\times 10^{-2}\,\mathrm{rad\,s^{-1}}$ in the calculation of Figure~\ref{fig:libero}(c).) prevents the system from reaching the desired final state, causing it instead to drift toward an expectation value of $\sigma_z$ corresponding to the $\ket{0}$ state. The pulse must therefore be optimized and extended in term of duration to obtain the correct final state. Subsequently, decoherence drives the system back toward $\ket{0}$ over time, limiting the operation features of its use as a qubit.

\begin{figure}[h!]
    \centering
    \includegraphics[width=0.45\textwidth]{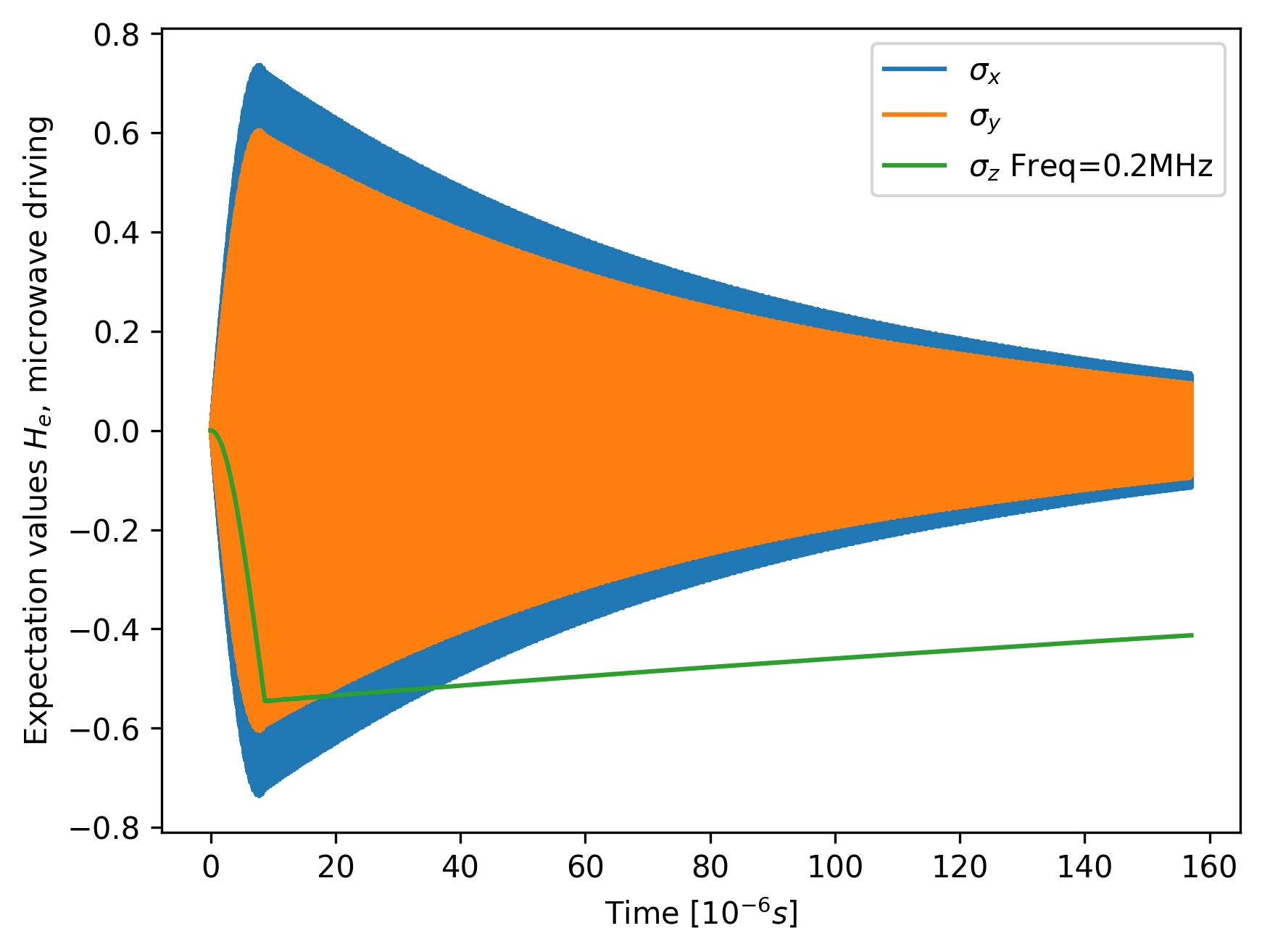}
    \includegraphics[width=0.45\textwidth]{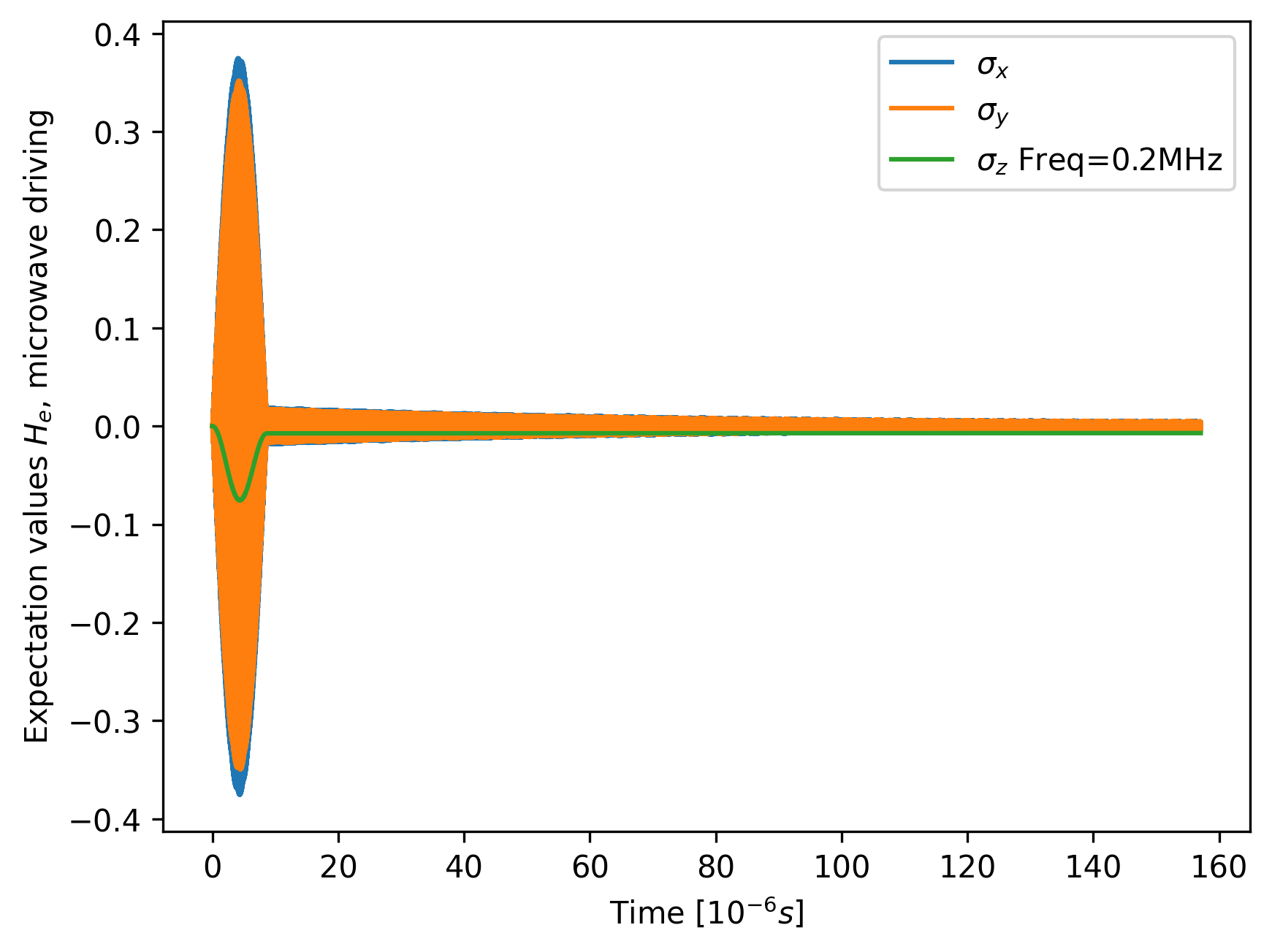}

    \vspace{0.3em} 
    \includegraphics[width=0.45\textwidth]{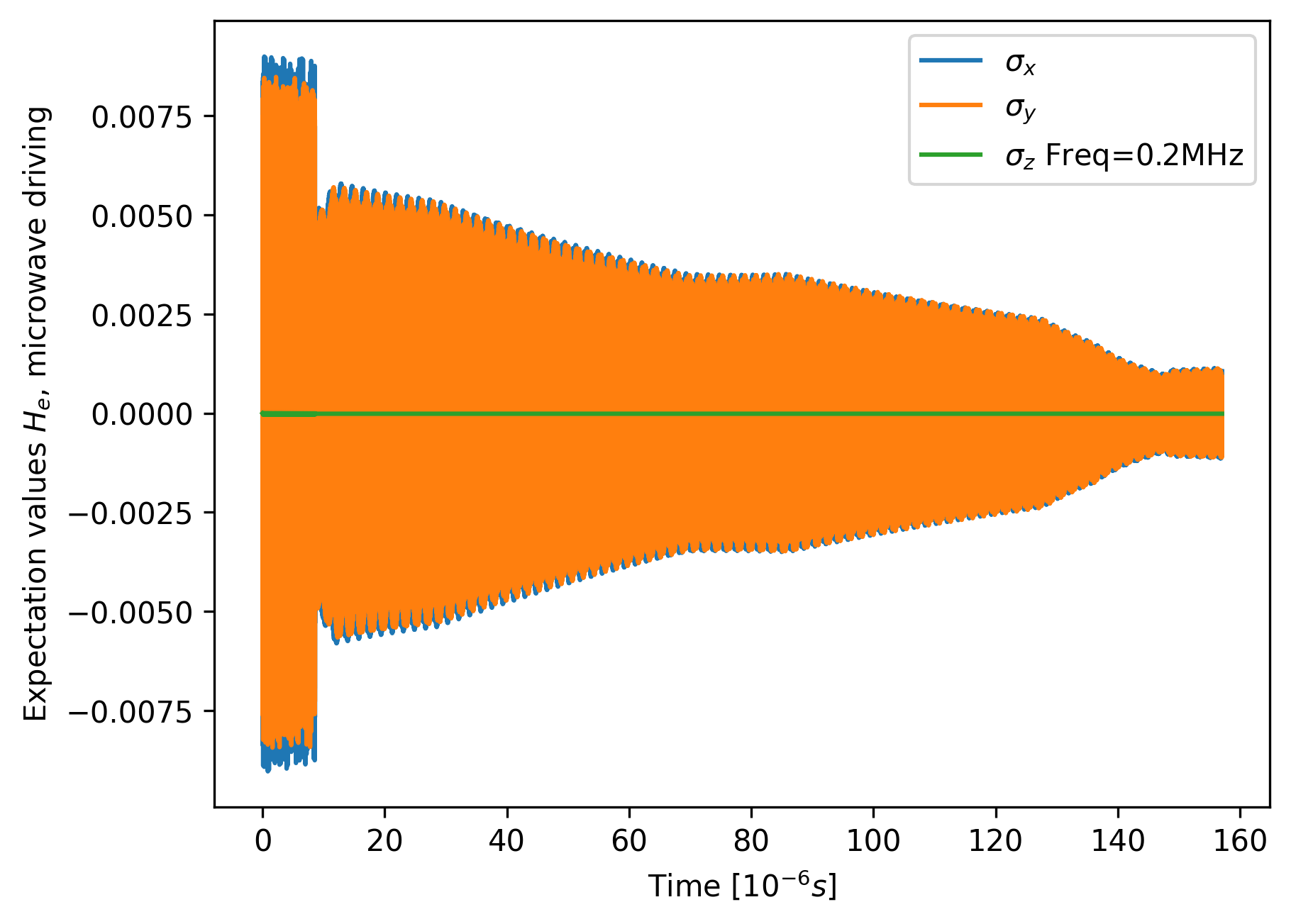}
    \includegraphics[width=0.45\textwidth]{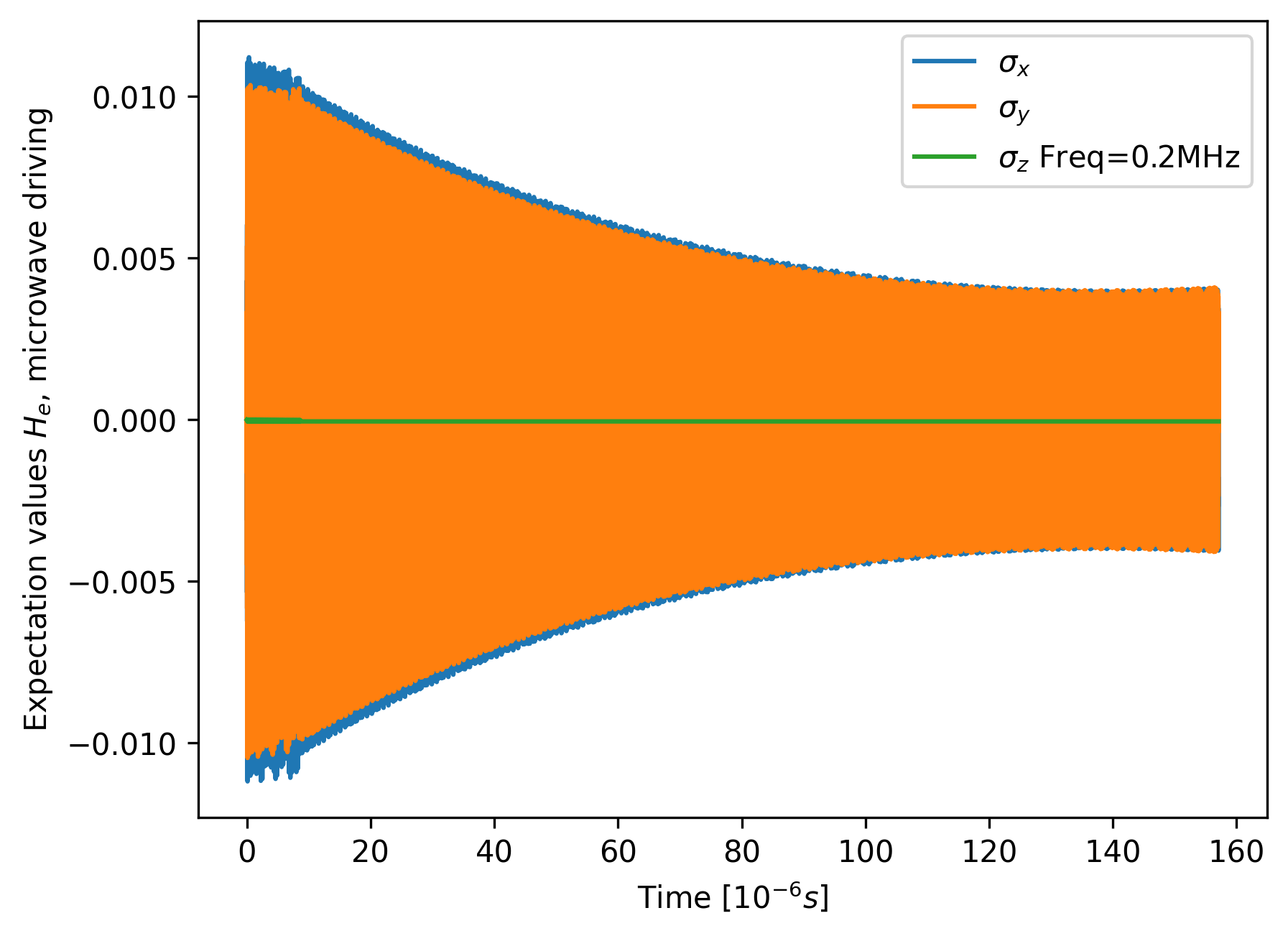}
    \vspace{0.1em}
    \caption{ Expectation values of the spin operators under a Rabi excitation pulse, tailored on the L6 configuration state energies, and dissipation. Configurations, starting from the upper left and proceeding clockwise, correspond to L6, L5, L4, and L3.}
    \label{fig:exicited}
\end{figure}

The drifting behavior is explicitly illustrated in Figure~\ref{fig:libero}(d). In this case, the simulation, together with the applied RF pulse, is ideally extended to the infinite-time limit. This reveals that, asymptotically, the decoherence terms dominate the system’s dynamics, leading to a natural drift of the expectation value toward $\braket{\sigma_z} = -0.25$.
A particularly relevant question concerns the degree of sensitivity of the dynamic state control process to variations in the Rabi excitation amplitude. This issue becomes significant in materials hosting multiple divacancies located at different depths, where local environments may modify the effective driving strength. In Figure~\ref{fig:exicited}, all axial divacancy configurations are subjected to RF pulses corresponding to a single $\frac{\pi}{2}$ rotation, applied with the same square-wave energy amplitude — specifically, the Rabi excitation value calculated for the L6 configuration. It is evident that, for all other cases, the target state, i.e.  is not achieved, highlighting the strong parameter dependence of the control dynamics and the lack of robustness of the excitation protocol across distinct divacancy configurations (a similar analysis is shown in the Supporting Information for the case of basal divacancies).

When a Rabi excitation is applied as a radiofrequency impulse with a square-modulated cosine waveform, its frequency is set according to the procedure described in the Computational Details for configuration L6. The duration of each impulse is chosen to correspond to a $\pi/2$ pulse at the Rabi frequency, i.e., $\tau = \frac{\pi}{2\Omega_R}$. Five consecutive pulses were applied for each configuration. We observe that the divacancy is highly sensitive to the excitation frequency: the L6-calculated Rabi excitation induces coherent dynamics in configuration L6 but is ineffective for the other configurations.

\section{Conclusion}
\justifying

This ab-initio based investigation clarifies how the structural and spin properties of neutral silicon–carbon divacancies in 3C–SiC are affected by their proximity to a hydrogen-terminated Si-rich (001) surface. The DFT results show that the defect geometry remains stable across all examined depths and orientations, with relaxations localized within the first coordination shells. The hydrogen passivation of the $(2 \times 1)$ surface effectively suppresses surface states within the band gap, thereby preventing unwanted electronic interference and preserving the intrinsic electronic characteristics of the divacancy even near the surface.
Spin-polarized calculations confirm that the ground state of all configurations is a spin triplet, consistent with bulk behavior. However, the zero-field-splitting (ZFS) parameters reveal clear surface-induced modifications. The axial component $D$ and the transverse component $E$ depend on the orientation of the defect and the distance from the surface. Basal configurations exhibit larger $D$ values than their axial counterparts and than those found in bulk, suggesting that $D$ may serve as an orientation marker for near-surface defects. In contrast, $E$ shows a stronger and more complex sensitivity to surface proximity: for the axial defects $|E|$ decreases rapidly with depth, while for basal configurations, its variation is smaller and non-monotonic in the investigated range of configurations, likely reflecting a less pronounced dependence of the defect related deformation field from the surface-defect distance.
Overall, these findings indicate that while the local electronic environment of the divacancy remains stable, symmetry breaking near the hydrogenated surface subtly alters its spin Hamiltonian. A defect exhibiting a single EPR signature in bulk thus gives rise to multiple spin-resolved features as it approaches the surface. This interplay between surface strain, structural relaxation, and spin-state fine structure highlights the importance of precise surface control in the engineering of SiC-based quantum systems.

Building on these static DFT results, we further explored the spin dynamics of near-surface divacancies through time-dependent simulations, examining their coherence behavior and the feasibility of spin-state manipulation under resonant radiofrequency (RF) excitation.
The state dynamics simulations indicate that the preparation of the divacancy electronic spin in mixed target states via an RF-based protocol is promising. These states are achievable even in an open system subject to decoherence, provided that slight adjustments are made to the pulse duration to compensate for the drift in the expectation values. Compared to previous studies, the timescales involved here are on the order of $10^{-4}\,\mathrm{s}$, whereas other simulations for bulk Si vacancies reported timescales of approximately $10^{-3}\,\mathrm{s}$.\cite{Fazio2024-il} The conventional protocol exhibits a high sensitivity to the depth and orientation of the divacancy, focusing its effectiveness in a single configuration for systems containing multiple divacancies. 

The coupled static and dynamic analyses presented in this work offer a comprehensive understanding of how surface proximity affects both the structural and quantum mechanical behavior of divacancy spin centers in 3C–SiC. From a broader perspective, these findings highlight the dual role of the surface: it acts both as a perturbing environment and as a potential tool for engineering spin properties through controlled symmetry breaking. The demonstrated robustness of the spin-triplet ground state and the retention of coherence under resonant excitation confirm that 3C–SiC remains a viable and tunable platform for quantum information technologies. Moreover, the pronounced sensitivity of the ZFS parameters to defect depth and orientation suggests that near-surface divacancies can serve as localized probes of strain and electrostatic fields, opening pathways for their use as nanoscale quantum sensors in solid-state devices.\\


\medskip
\textbf{Supporting Information} \par 
Supporting Information is available from the Wiley Online Library or from the author.

\medskip
\textbf{Acknowledgements} \par 
\justifying
The work was partially funded by the Italian Ministry University and Research (MUR) in the framework of project PNRR Partenariato PE4 NQSTI Quantum, Grant No. PE0000023. I.D and G.F. acknowledge the funding of Ministero delle Imprese e del Made in Italy (MIMIT) in the framework of the Important Project of Common European Interest on Microelectronics and Communication Technologies, IPCEI ME/CT, Grant No. SA.101186.
Open access publishing facilitated by Consiglio Nazionale delle Ricerche, as part of the Wiley - CRUI-CARE agreement.

\medskip
\textbf{Conflict of interest} \par
The authors declare no conflict of interest.
\medskip

\begin{figure}[h!]
\textbf{Table of Contents}\\
Near-surface divacancy defects in cubic silicon carbide exhibit modified properties due to surface-induced symmetry breaking, as revealed by first-principles calculations. The spin Hamiltonian is modulated by the defect’s depth and orientation, providing key insights for engineering robust SiC-based quantum devices and nanoscale sensors.

\medskip
\Centering
  \includegraphics{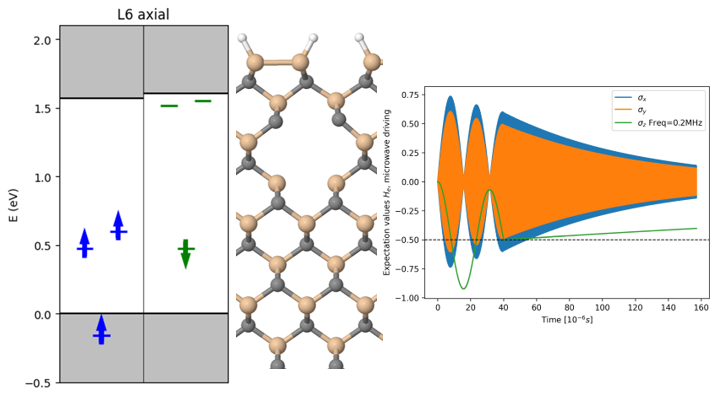}
  \medskip
  \caption*{ToC Entry}
\end{figure}




\begin{thebibliography}{10}

\bibitem{LarsJaeger}
Lars Jaeger.
\newblock {\em The Second Quantum Revolution: From Entanglement to Quantum Computing and Other Super-Technologies}.
\newblock Springer, Switzerland, 2018.

\bibitem{Awschalom2018}
D.~D. Awschalom, R.~Hanson, J.~Wrachtrup, and B.~B. Zhou.
\newblock Quantum technologies with optically interfaced solid-state spins.
\newblock {\em Nature Photonics}, 12(9):516--527, 2018.

\bibitem{Castelletto2020-qd}
S.~Castelletto and A.~Boretti.
\newblock Silicon carbide color centers for quantum applications.
\newblock {\em Journal Physics: Photonics}, 2(2):022001, 2020.

\bibitem{Seo2016-vy}
S.~Hosung, A.~L. Falk, P.~V. Klimov, V.~Paul, K.~Miao, G.~Galli, and D.~D. Awschalom.
\newblock Quantum decoherence dynamics of divacancy spins in silicon carbide.
\newblock {\em Nature Communications}, 7(1):12935, 2016.

\bibitem{Fazio2024-il}
T.~Fazio, I.~Deretzis, G.~Fisicaro, E.~Paladino, and A.~La Magna.
\newblock Stability and decoherence analysis of the silicon vacancy in {3C-SiC}.
\newblock {\em Physical Review A (Coll. Park.)}, 109(2), 2024.

\bibitem{Falk2014}
A.~L. Falk, P.~V. Klimov, B.~B. Buckley, et~al.
\newblock Polytype control of spin qubits in silicon carbide.
\newblock {\em Physical Review Letters}, 112(18):187601, 2014.

\bibitem{Koehl2011}
W.~F. Koehl, B.~B. Buckley, F.~J. Heremans, G.~Calusine, and D.~D. Awschalom.
\newblock Room temperature coherent control of defect spin qubits in silicon carbide.
\newblock {\em Nature}, 479(7371):84--87, 2011.

\bibitem{Xuan2025}
Z.H. He, J.Y Zhou, Q.~Li, et~al.
\newblock Robust single modified divacancy color centers in 4h-sic under resonant excitation.
\newblock {\em Nature Communications}, 15:10146, 2024.

\bibitem{Zhu2023}
Y.~Zhu, V.~W.Z. Yu, and G.~Galli.
\newblock First-principles investigation of near-surface divacancies in silicon carbide.
\newblock {\em Nano Letters}, 23(22):11453--11460, 2023.

\bibitem{Lukin2023}
D.M. Lukin, M.A. Guidry, J.~Yang, et~al.
\newblock Two-emitter multimode cavity quantum electrodynamics in thin-film silicon carbide photonics.
\newblock {\em Physical Review X}, 13(1):011005, 2023.

\bibitem{JiYangZhou2023}
J.-Y. Zhou, Q.~Li, Z.H. Hao, et~al.
\newblock Plasmonic-enhanced bright single spin defects in silicon carbide membranes.
\newblock {\em Nano Letters}, 23(10):4334--4343, 2023.

\bibitem{Calogero2025}
G.~Calogero, I.~Deretzis, G.~Fisicaro, D.~Ricciarelli, R.~G. Viglione, and A.~La Magna.
\newblock Tailoring nuclear spins order with defects: A quantum technology cad study.
\newblock {\em Advanced Quantum Technologies}, 8:e2500160--e2500172, 2025.

\bibitem{KitsonPhysRevA2024}
P.~Kitson, T.~Haug, A.~La Magna, O.~Morsch, and L.~Amico.
\newblock Rydberg atomtronic devices.
\newblock {\em Phys. Rev. A}, 110:043304, 2024.

\bibitem{Hamann2013}
D.~R. Hamann.
\newblock Optimized norm-conserving vanderbilt pseudopotentials.
\newblock {\em Physical Review B}, 88(8):085117, 2013.

\bibitem{Giannozzi2009}
P.~Giannozzi, S.~Baroni, N.~Bonini, et~al.
\newblock Quantum espresso: a modular and open-source software project for quantum simulations of materials.
\newblock {\em Journal of Physics: Condensed Matter}, 21(39):395502, 2009.

\bibitem{PBE1996}
J.~P. Perdew, K.~Burke, and M.~Ernzerhof.
\newblock Generalized gradient approximation made simple.
\newblock {\em Physical Review Letters}, 77(18):3865--3868, 1996.

\bibitem{Rayson2008}
M.~J. Rayson and P.~R. Briddon.
\newblock First principles method for the calculation of zero-field splitting.
\newblock {\em Physical Review B}, 77(3):035119, 2008.

\bibitem{pyzfs}
H.~Ma, M.~Govoni, and G.~Galli.
\newblock Pyzfs: A python package for first-principles calculations of zero-field splitting tensors.
\newblock {\em Journal of Open Source Software}, 5(47):2160, 2020.

\bibitem{Paw}
A.~Dal Corso.
\newblock Pseudopotentials periodic table: From h to pu.
\newblock {\em Computational Materials Science}, 95:337 – 350, 2014.

\bibitem{lambert2024qutip5quantumtoolbox}
N.~Lambert, E.~Giguère, P.~Menczel, et~al.
\newblock Qutip 5: The quantum toolbox in python.
\newblock 2024.

\bibitem{Christle}
D.~J. Christle, P.~V. Klimov, C.~F. de~las Casas, et~al.
\newblock Isolated spin qubits in sic with a high-fidelity infrared spin-to-photon interface.
\newblock {\em Physical Review X}, 7:021046, 2017.

\bibitem{Cai2013-bc}
J.~Cai, A.~Retzker, F.~Jelezko, and M.~B. Plenio.
\newblock A large-scale quantum simulator on a diamond surface at room temperature.
\newblock {\em Nat. Phys.}, 9(3):168--173, 2013.

\bibitem{Zaiser2016-va}
S.~Zaiser, T.~Rendler, I.~Jakobi, et~al.
\newblock Enhancing quantum sensing sensitivity by a quantum memory.
\newblock {\em Nat. Commun.}, 7(1):12279, 2016.

\end{thebibliography}
\end{document}